# Low work function thin film growth for high efficiency thermionic energy converter: coupled Kelvin probe and photoemission study of potassium oxide

F. Morini [*, 1], E. Dubois [1], J. F. Robillard [1], S. Monfray [2], T. Skotnicki [2]

[1] Institut d'électronique, de microélectronique et des nanotechnologies (IEMN), UMR CNRS 8520, Cité Scientifique, Avenue Poincaré, BP 60069, 59652 Villeneuve d'Ascq Cedex, France
[2] ST Microelectronics, 850, rue Jean Monnet, 38926 Crolles, France,

Recent researches in thermal energy harvesting have revealed the remarkable efficiency of thermionic energy converters comprising very low work function electrodes. From room temperature and above, this kind of converter could supply low power devices such as autonomous sensor networks. In this type of thermoelectric converters, current injection is mainly governed by a mechanism of thermionic emission at the hot electrode which explains the interest for low work function coating materials. In particular, alkali metal oxides have been identified as excellent candidates for coating converter electrodes. This paper is devoted to the synthesis and characterization of potassium peroxide $K_2O_2$ onto silicon surfaces. To determine optimal synthesis conditions of $K_2O_2$, we present diagrams showing the different oxides as a function of temperature and oxygen pressure from which phase stability characteristics can be determined. From the experimental standpoint, we present results on the synthesis of potassium oxide under ultra high vacuum and controlled temperature. The resulting surface is characterized *in-situ* by means of photoemission spectroscopy (PES) and contact potential difference (CPD) measurements. A workfunction of 1.35 eV is measured which and the expected efficiency of the corresponding converter is discussed. It is generally assumed that the decrease of the work function in the alkali/oxygen/silicon system, is attributed to the creation of a surface dipole resulting from a charge transfer between the alkali metal and oxygen.

**1 Introduction** Thermionic energy converters (TEC) are made of two parallel electrodes separated by a vacuum gap. Thermionic emission at the electrodes surfaces increases with temperature such that a temperature gradient results in a net current flow across the gap [1]. The principles of these converters radically contradict conventional solid state thermoelectric generation based on the Seebeck effect. Firstly, electronic transport occurs in the ballistic regime between the electrodes. Secondly, the main heat transfer process is radiative as compared to conductive in conventional generators. The subsequent dramatic reduction of heat losses leads to an increased efficiency. Finally, maximizing efficiency relies essentially on minimizing the electrodes workfunction and adjusting the gap width [2]. By comparison, thermoelectric materials optimization involves difficulties due to the antagonisms between the key transport parameters, namely, the Seebeck coefficient, the electrical conductivity and the thermal conductivity.

The thermionic current is governed by the Richardson equation and is therefore extremely sensitive to the temperature and electrode material work function [3]:

$$J = A T^2 \exp\left(-\frac{q(\phi + V)}{kT}\right) \quad (1)$$

where J is the thermionic current density (A.cm$^{-2}$), A the Richardson constant (120 A.cm$^{-2}$.T$^{-2}$), $\phi$ the material work function (eV), q the elementary charge, V the applied bias, k the Boltzmann constant and T the absolute temperature. Since electrodes are often constituted of metal and semiconductors with work function typically ranging from 3eV to 6eV, the temperature window for efficient thermionic conversion typically lies in the 600-1000K interval [4].

Nowadays fast developments in nanotechnology enable thin film coating of metal electrode surface [5]. Recent studies have identified potassium oxide as one of best candidates to decrease the work function of classical semiconductors therefore to enabling lower TEC operating temperature or increased efficiency [6,7].

We present in a first part the calculated efficiency of an ideal TEC and its associated power output density as a function of the electrode work function for a range of temperatures. In a second part, volatility diagram of the potassium

* Corresponding author: e-mail francois.morini@isen.iemn.univ-lille1.fr, Phone: +33 320 197 917 , Fax: +33 320 197 884

oxygen binary system are presented to target the synthesis conditions of potassium oxide thin film. In the last part, work function measurements at room temperature are discussed based on Kelvin probe and photoemission techniques.

## 2 Thermionic energy converter efficiency

A TEC schematic is given in figure 1. The hot (emitter) and cold (collector) electrodes are connected by an external resistive load.

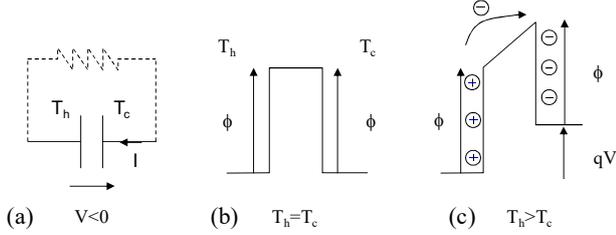

**Figure 1** (a) TEC electrical representation, (b) & (c) electron potential energy in space. $T_H$ and $T_C$ are respectively the temperature of the hot side and the cold side.

We evaluated maximum efficiency of a TEC following assuming that the two electrodes feature the same work function. Details about the calculation can be found elsewhere [8]. The calculation takes into account the net current density is given by the difference between the emitter and the collector. In this ideal case, thermal losses are reduced to the heat carried by electron from the emitter to the collector and by radiative transfer which is independent of the distance between electrodes and follows the Stephan-Boltzmann's law. Figure 2 clearly shows the interest of lowering the electrode work function. It is shown that conversion efficiency as high as 20% can be reached even with a 100K temperature gradient provided the material work function is less than 0.5eV. To obtain such performance in thermopiles converters would require materials featuring very high zT.

## 3 Material volatility diagrams

To determine optimal synthesis conditions of $K_2O_2$, we established the volatility diagram at 1000K as a function of potassium vapor pressure and oxygen pressure from which phase stability can be determined. From thermochemical data [9-12], $K_2O$, $K_2O_2$ and $KO_2$ phases are considered for state functions calculations as a function of oxygen partial pressure.

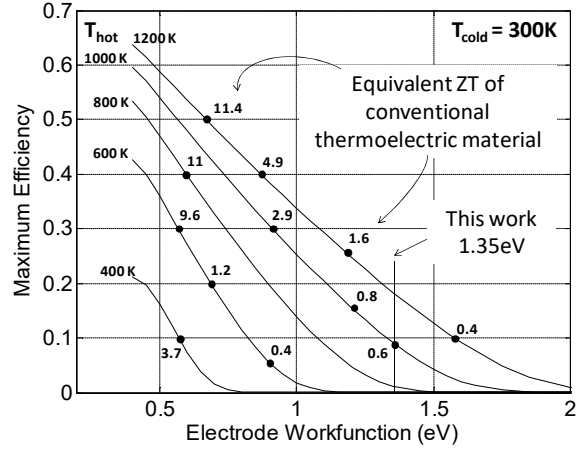

**Figure 2** Ideal TEC efficiency as a function of the electrode work function. The emitter temperature is set to 400K and increases by step of 200K up to 1600K. The collector temperature is fixed to 300K. The theoretical efficiencies of various solid state thermoelectric generators based on the material figure of merit (zT) are indicated. The workfunction obtained in this study is shown as a vertical line.

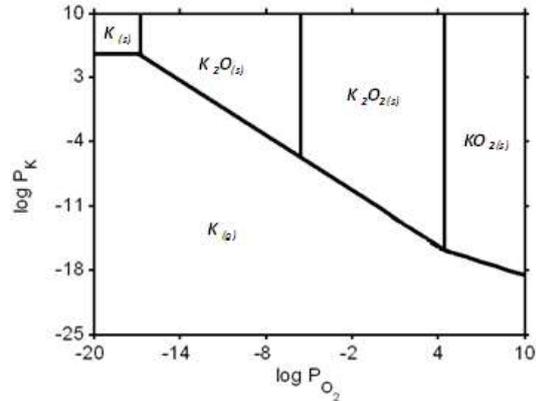

**Figure 3** Volatility diagram of the potassium oxygen binary system at 1000K.

Table 1 summarizes the equilibrium constants χ associated to each potassium oxidation step calculated from Eq. 2:

$$\chi_i = -\frac{\Delta_R G_i(T)}{RT} = -\frac{\Delta_R H_i(T)}{RT} + \frac{\Delta_R S_i(T)}{R}, \quad (2)$$

where $\Delta_R G_i(T)$, $\Delta_R H_i(T)$, $\Delta_R S_i(T)$ and R are respectively the free Gibbs energy, the enthalpy of the reaction, the entropy of the reaction and the molar gas constant. Figure 3 shows the synthesis domain of potassium oxides and brings information on the pressure range in order to synthesize the desired phase.

**Table 1** Thermo chemical parameters used for volatility diagram presentation at 1000K.

| Reaction | Eq. constant χ at T | Equation of the equilibrium line | Log χ at T * |
|---|---|---|---|
| $K_{(g)} = K_{(s)}$ | $\chi_1 = P°·P_K^{-1}$ | $\text{Log } P_K = 5 - \text{Log } \chi_1$ | $\text{Log } \chi_1 = -0.499$ |
| $4 K_{(g)} + O_{2(g)} = 2 K_2O_{(s)}$ | $\chi_2 = P°·P_K^{-1}·P_{O2}^{-1}$ | $\text{Log } P_K = 0.25 (5 - \text{Log } P_{O2} - \text{Log } \chi_2)$ | $\text{Log } \chi_2 = 54.2$ |
| $2\,K_{(g)} + O_{2\,(g)} = K_2O_{2\,(s)}$ | $\chi_3 = P°.P_K^{-2}.P_{O2}^{-1}$ | $\text{Log } P_K = 0.5\,(5 - \text{Log } P_{O2} - \text{Log } \chi_3)$ | $\text{Log } \chi_3 = 32.1$ |
| $K_{(g)} + O_{2\,(g)} = KO_{2\,(s)}$ | $\chi_4 = P°.P_K^{-1}.P_{O2}^{-1}$ | $\text{Log } P_K = (5 - \text{Log } P_{O2} - \text{Log } \chi_4)$ | $\text{Log } \chi_4 = 16.3$ |
| $4K_{(s)} + O_{2\,(g)} = 2\,K_2O_{(s)}$ | $\chi_5 = P°.P_{O2}^{-1}$ | $\text{Log } P_{O2} = 5 - \text{Log } \chi_5$ | $\text{Log } \chi_5 = 56.2$ |
| $2\,K_2O_{(s)} + O_2 = 2\,K_2O_{2\,(s)}$ | $\chi_6 = P°.P_{O2}^{-1}$ | $\text{Log } P_{O2} = 5 - \text{Log } \chi_6$ | $\text{Log } \chi_6 = 10.0$ |
| $K_2O_{2\,(s)} + O_2 = 2\,KO_{2\,(s)}$ | $\chi_7 = P°.P_{O2}^{-1}$ | $\text{Log } P_{O2} = 5 - \text{Log } \chi_7$ | $\text{Log } \chi_7 = 0.489$ |

\* $\text{Log } \chi_i$ is calculated using the numeric value of free Gibbs energy at the equilibrium state for each reaction

**4 Experimental** The p-doped Si (100) sample (5–10 Ω.cm) was oxidized and cleaned using a hydrogen peroxide and sulfuric acid mixture ($H_2SO_4 : H_2O_2$ / 1:1 in volume) for 20 min. followed by a de-oxidation step in hydrofluoric acid (HF 5%) to get a H-terminated silicon (Si) surface. This process sequence is known to occupy Si dangling bonds as already verified by contact angle, XPS and STM techniques [13]. Potassium was deposited onto the H:Si(100) substrate at room temperature using a well-outgased SAES dispenser with a 7.5A current intensity under an oxygen atmosphere for 20 min. The potassium gas temperature is estimated to 1000K. The gas pressure was set at $3\times10^{-7}$ mbar and held constant during K deposition. Kelvin probe measurements were performed under a base pressure of $3\times10^{-10}$ mbar at room temperature to obtain the contact potential difference (CPD) between the surface and the vibrating tip. For that sake, a KP Technology apparatus with a 2 mm diameter circular stainless steel tip has been used. Complementarily, absolute work functions have been measured using photoemission spectroscopy (PES) under monochromatic illumination ranging from 1.7eV and 3.5eV.

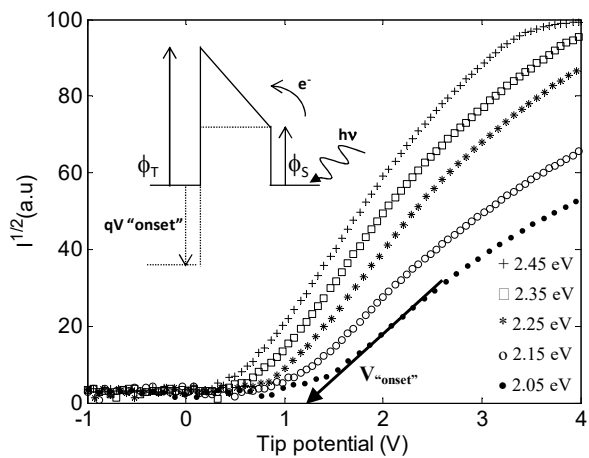

**Figure 4** I/ V representation in photocurrent measurement mode including an energy level diagram showing the influence of the tip potential V on the photocurrent detection. $\phi_T$ and $\phi_S$ are respectively the tip and sample work functions.

**5 Discussion** To determine the tip work function, the sample is lighted at fixed photon energy hν to trig photoelectron emission. From the energy band diagram, it is necessary to apply a tip voltage for photoelectron detection to shift the tip barrier height below the sample one. The "onset" voltage is determined when photocurrent is measured to the tip. The operation is made for a large photon energy range. The results are presented on the figure 4. As expected, the "onset" voltage is linearly decreasing with the photon energy as shown on figure 5 and the tip work function is determined at $V_{onset}=0$. The sample workfunction is also determined by PES. Figure 6 shows the photocurrent obtained as a function of the photons energy for a high tip potential (10V) to shift the tip Fermi level below the sample Fermi level. The photon energy range of this experiment is above the sample work function.

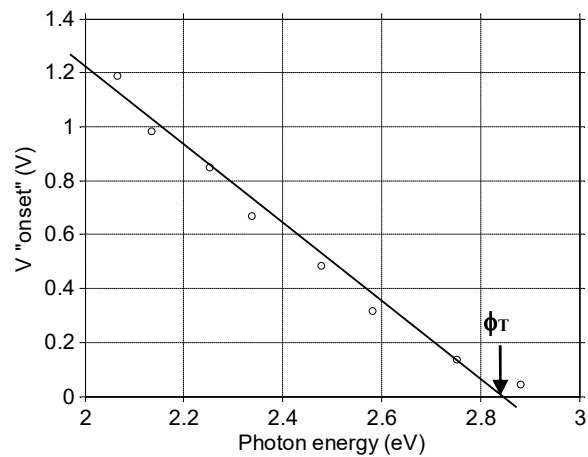

**Figure 5** Onset voltage of the photocurrent ($V_{"onset"}$) as a function of the photon energy.

These results are supported by CPD measurements of the Si-H surface before and after potassium oxide thin-film deposition (Figure 6 inset). This graph also shows the excellent stability of the surface properties and of the measurement repeatability over a typical time of 6 min. The CPD is a relative measure of sample work function with respect to the tip. Thus, the inset on figure 6 shows a relative decrease of the surface work function by 1.49eV.

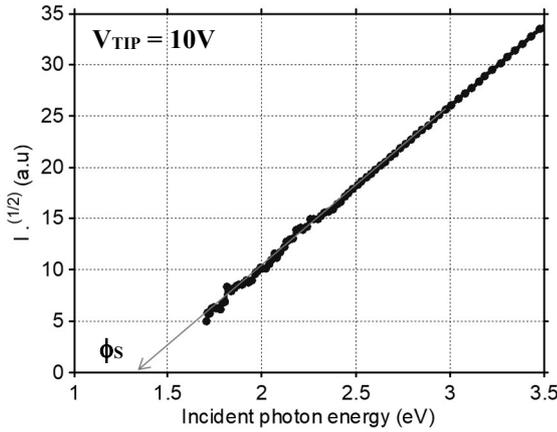

**Figure 6** Plot of the square root of photo current ($I^{1/2}$) as a function of incident photon energy hν in the visible range. Inset: Plot of the CPD before and after potassium oxide deposition. Each data point is the average of 70 measurements.

Essential results, summarized in table 2, indicate that the absolute work function of the potassium oxide surface is lowered down to 1.35eV while bare silicon surfaces feature a typical work function in the 4.1-5.2eV range depending on doping, orientation and surface treatment. When compared to metallic potassium coated surfaces for which a typical work function of 2.3eV is reported in the literature [15-16], the low bound of 1.35eV also clearly indicates that potassium has reacted with oxygen. A TEC based on these result and operating between $T_c$=300K and $T_h$=1000K value would achieve a 10% efficiency corresponding to a zT=0.6 material.

**Table 2** Work functions of potassium oxide coated sample and tip from CPD and PES experiments.

| Material | CPD (V) | Work function (eV) |
|---|---|---|
| K/O: Si (100) | -1.49 | 1.35 |
| Tip | | 2.85 |

We now discuss the origin of the lower workfunction in terms of dipole moment formation. Due to the oxygen high electronegativity value (Pauling's: 3.5eV), a dipole moment μ is likely created between potassium and oxygen [17]. The dipole moment created at the surface between potassium and oxygen is given by:

$$\mu_{K-O} = d_{K-O} \cdot \delta_{K-O} \cdot q, \quad (3)$$

where $d_{K-O}$ is the length of the potassium oxygen bond, $\delta_{K-O}$ stands for the percent ionic character of the single chemical bond and q is the elementary charge. The percent ionic character and the dipole concentration are respectively given by the Eq. 4 and Eq.5:

$$\delta_{K-O}(\%) = 100 \times \left[1 - \exp\left(-\frac{(\chi_K - \chi_O)^2}{4}\right)\right] \quad (4)$$

$$\Delta\phi = \frac{q \cdot \mu \cdot N_{K-O}}{\varepsilon_0}, \quad (5)$$

with, χ the Pauling's electronegativities of the considered elements, $N_{K-O}$ the number of dipole per surface unit and $\varepsilon_0$ the permittivity of vacuum [17-19]. According to this approach, the percent ionic character of the potassium oxygen bond amounts to 84% with a bonding length of 2.73Å. The estimated dipole moment and concentration are estimated at 10.1D and $8\times10^{13}$ cm$^{-2}$, respectively.

**6 Conclusion** Low work function coatings have been synthesised by evaporating potassium in oxidant atmosphere on hydrogen passivated (100) p-type silicon at room temperature. Using CPD and photoemission techniques, work function values have been measured before and after deposition and are found to be equal to 4.70eV and 1.35eV, respectively. Assuming that the K-O compound is adsorbed on the silicon surface, the important decrease of the work function by 3.35eV can be explained by a surface dipole model [17].